\documentclass[12pt]{iopart}
\usepackage{iopams}
\usepackage{graphicx}
\usepackage{mathtext}

\begin{document}

\newcommand {\eps} {\varepsilon}
\newcommand {\Nt} {{\mbox{\scriptsize N}}}
\newcommand {\Ct} {{\mbox{\scriptsize C}}}
\renewcommand {\Re} {\mathrm{Re}}
\renewcommand {\Im} {\mathrm{Im}}
\newcommand {\la} {\langle}
\newcommand {\ra} {\rangle}
\newcommand {\lla} {\left\langle}
\newcommand {\rra} {\right\rangle}

\title[Towards prediction of proteasomal digestion patterns of proteins]
{Towards quantitative prediction of
 proteasomal digestion patterns of proteins}

\author{Denis S Goldobin$^{1,2}$ and Alexey Zaikin$^{3,4}$}
\address{$^1$Department of Physics, University of Potsdam,
        Postfach 601553, D--14415 Potsdam, Germany}
\address{$^2$Department of Theoretical Physics, Perm State University,
        15 Bukireva str., 614990 Perm, Russia}
\address{$^3$Department of Mathematical \& Biological Sciences, University of Essex,
        Wivenhoe park, CO4~3SQ Colchester, UK}
\address{$^4$Departments of Mathematics \& Institute of Women Health,  University College London,
        Gower street, WC1E~6BT London, UK}
\ead{Denis.Goldobin@gmail.com}

\begin{abstract}
We discuss the problem of proteasomal degradation of proteins.
Though proteasomes are important for all aspects of the cellular
metabolism, some details of the physical mechanism of the process
remain unknown. We introduce a stochastic model of the proteasomal
degradation of proteins, which accounts for the protein
translocation and the topology of the positioning of cleavage
centers of a proteasome from first principles. For this model we
develop the mathematical description based on a master-equation
and techniques for reconstruction of the cleavage specificity
inherent to proteins and the proteasomal translocation rates,
which are a property of the proteasome specie, from mass
spectroscopy data on digestion patterns. With these properties
determined, one can quantitatively predict digestion patterns for
new experimental set-ups. Additionally we design an experimental
set-up for a synthetic polypeptide with a periodic sequence of
amino acids, which enables especially reliable determination of
translocation rates.
\end{abstract}

\vspace{2pc}
\noindent{\it Special Issue}: Article preparation, IOP journals
\maketitle


A macromolecular complex, the proteasome, is the complex molecular
machine for the degradation of intracellular
proteins~\cite{1994_Rock}. In particular, proteasomes  produce epitopes for an immune
system~\cite{2001_Kloetzel_nature}. They exist in cells as the
free proteolytically active core, the barrel-shaped 20S proteasome
(\fref{fig1}), and as associations of this core with regulatory
complexes PA700 (19S regulator) or PA28 (11S regulator) at its
ends~\cite{2000_Tanahashi}. This paper deals with proteasomal
digestion of proteins widely studied in molecular biology and
immunology.

A protein enters the proteasome and is transported into the
central chamber (this process is referred as the {\it
translocation} one) where it is cleaved into fragments by one of
the cleavage terminals arranged along two rings. Fragments of the
protein produced are removed through proteasome gates. Some of
these fragments, epitopes, are transported onto the cell surface
where T-lymphocytes scan them in order to recognize the cells to
be killed because of an abnormal functioning. Hence, the digestion
pattern for a degraded protein and its statistical properties
determine the reaction of the immune system to the presence of
this protein in a certain cell. Peculiarities of the translocation
rates can qualitatively affect the expression of the specific
fragment, {\it e.g.}, an epitope, because an altered transport
changes time of being near the cleavage terminal, {\it i.e.},
conditions of cleavage. Moreover, impairment of proteasomal
degradation, probably due to transport malfunction, might
contribute to the pathology of various neurodegenerative
conditions~\cite{2006_Rubinsztein_nature}.

The mechanism of protein translocation remains unknown (however,
subjects related to some extent to this problem have been studied
in~\cite{2000_Holzhuetter_bj,2002_Peters,2005_Luciani,2005_Zaikin_proteasome}).
It is also unknown whether this mechanism is qualitatively
different for different proteasome types (constitutive or
immuno-), with/without different regulatory complexes. Recently,
in~\cite{2008_Goldobin_etal} a stochastic model, which allows a
straightforward reconstruction of the translocation rates and
cleavage specificities from mass spectroscopy (MS) data on
digestion patterns, has been introduced. These properties
reconstructed can be used for a comprehensive quantitative
prediction of proteasomal digestion patterns for new proteins and
new experimental set-ups. In this paper we elaborate the
mathematical theories for the employing of the introduced model
for relatively {\it short synthetic polypeptides} (\sref{sec2}),
{\it long proteins with a periodic sequence of amino acids}
(\sref{sec3}), and {\it long natural proteins} which require a
peculiar approach (\sref{sec4}).


\section{Physical model of the system and mathematical description}
\label{sec1}
\begin{figure}[!t]
\center{
  \includegraphics[width=0.97\textwidth]%
 {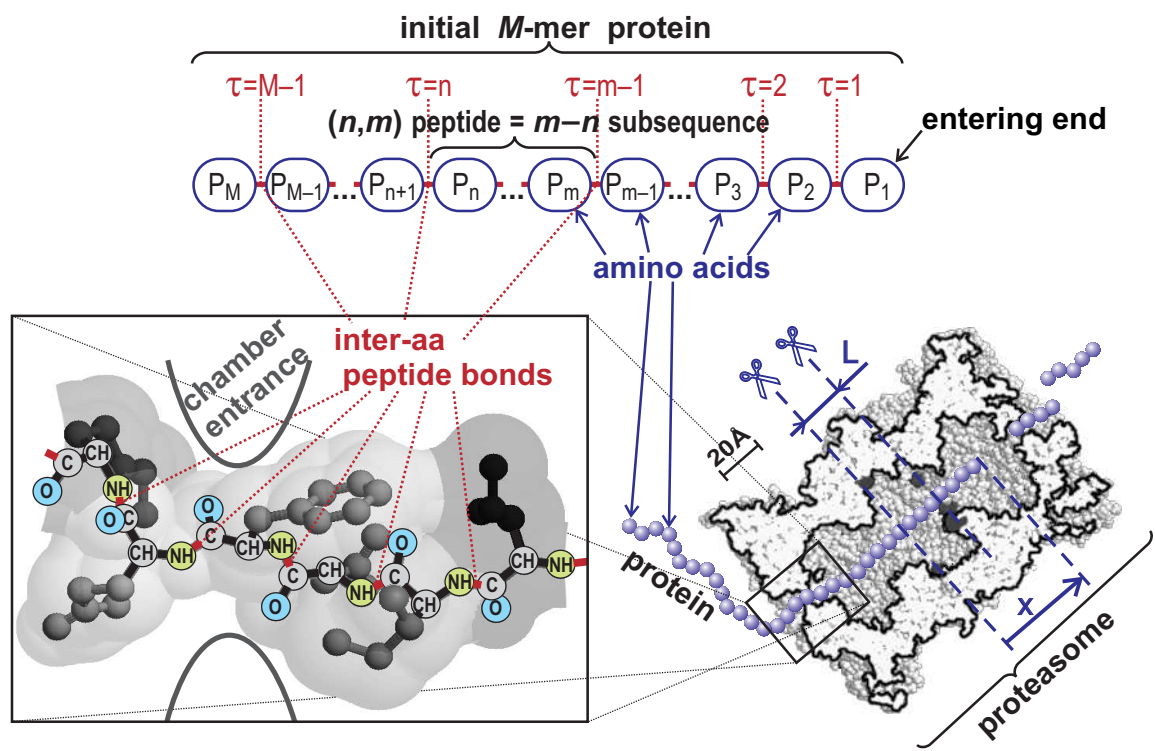}
}
  \caption{Infiltration of a protein strand into the 20S proteasome:
The scissors mark the positions of active sites rings at $x=0$ and
$x=L$; the cleavage occurs via the attaching-detaching of the
protein to active sites (dark-grey color). The zoom-in shows the
protein fragment {\tt KEFNII} passing through the gate; the
electron shields are presented in pale colors.}
  \label{fig1}
\end{figure}

We describe the process of protein transport and degradation by
the proteasome (see \fref{fig1}) within the framework of the
following assumptions.

\noindent
{\bf $\bullet$\;Protein translocation:} The process of the
infiltration of a protein into the proteasome chamber is a
sequence of thermal noise induced jumps of the protein strand by
one amino acid (AA). In \fref{fig1}, the zoom-in of the chamber
gate schematically shows the diameter of the gate to be comparable
with the characteristic size of an AA, what means that the protein
chain may be fixed in metastable states by a tight gate between
successive jumps due to large thermal fluctuations. Indeed, the
atomic force measurements reveals $U_b/kT\!>\!3$~\cite{2007_Witt},
where $U_b$ is the characteristic height of the energy barrier
separating nearest metastable positions of the chain and $kT/2$ is
the energy of thermal fluctuations. The probability of the protein
shift by one AA during the infinitesimal time interval $\rmd t$
into the proteasome (to the right in
\fref{fig1}) is assumed to depend only on the length $x$ of the
protein forward end beyond the active sites nearest to the
proteasome chamber gate used for protein infiltration (the left
ones in \fref{fig1}); this probability divided by $\rmd t$ is
given by the translocation rate function (TRF) $v(x)\equiv v_x$.
In such a way, we neglect the role of the AA sequence specificity
for translocation, what is suggested by a non-covalent interaction
between the proteasome and the retracted protein. The backward
motions of the entering strand are neglected as well
(from~\cite{2007_Witt}, for the potential energy $U(x)$ of the
metastable state $x$, $(U(x-1)-U(x+1))/2kT\approx2.5$, thus,
meaning the probability of a backward motion to be diminished by
factor $\e^{-2.5}$ against the forward one). These assumptions do
not impose significant restrictions on the physical mechanism of
the translocation process: they are valid for the thermal drift in
a tilted spatially-periodic potential ({\it e.g.},
see~\cite{2001_Reimann}) as well as for the ratchet effect ({\it
e.g.}, see~\cite{2005_Zaikin_proteasome}), {\it etc}. The TRFs of
different proteasome species (20S, 26S which is the association of
20S core and 19S regulatory complexes, {\em
etc.}~\cite{2000_Tanahashi}) differ.

\noindent
{\bf $\bullet$\;Cleavage:} When the protein strand is close to the
active site, the probability of cleavage during the infinitesimal
time interval $\rmd t$ depends on the sequence of AAs nearest to
the peptide bond cleaved~\cite{2005_Tenzer}. For the given
protein, this conditional probability divided by $\rmd t$, in
other words, conditional cleavage rate (CCR),
$\gamma(\tau)\equiv\gamma_\tau$, is a function of the bond number
$\tau$ (precisely, $\tau$ numerates the position of the bond
within the initial protein and is counted from the end which has
first entered the proteasome; see
\fref{fig1}). In the following we use the number $\tau$ of the
bong nearest to the first ring of active sites as a {\em time-like
variable}.

\noindent
{\bf $\bullet$\;Removal of digestion products:} The cleaved parts
of the protein degraded, peptides, leave the chamber through the
second proteasome gate. Due to their mobility being higher in
comparison to that of the protein, processed peptides leave the
chamber quick enough to neglect both their possible further
splitting and their influence on the protein translocation.

Let us now introduce the distribution $w(x|\tau)$ which is the
probability of the protein forward end beyond the first ring of
the active sites to be of the length $x$, when the $\tau$th bond
is near that ring, in terms we use henceforth, at the discrete
``time moment'' $\tau$. We measure $x$ in AA. Note, $x$ and $\tau$
are integer. In the following we describe the ``temporal''
evolution of distribution $w(x|\tau)$. On this way, we treat the
shift of the protein strand into the proteasome for one AA, {\em
i.e.}, the transition $\tau\to\tau+1$. Let us decompose
$w(x|\tau+1)$ as
\[\textstyle
w(x|\tau+1)=\sum_j w_j(x|\tau+1)\,,
\]
where $w_j(x|\tau+1)$ are the contributions due to different
scenarios of this transition. Along with $w(x|\tau)$, we account
$Q(n,m|\tau)$, the amount of the peptide $(n,m)$, which is the
$m$--$n$ subsequence of the degraded protein (see \fref{fig1}),
generated during transition $\tau\to\tau+1$.

In the process of protein digestion there are three possible elementary events:
\\
{\sf (a)}\;the strand shift: $x\to x+1$, $\tau\to\tau+1$; the
event rate is $v(x)$;
\\
{\sf (b)}\;the cleavage on the first ring of cleavage centers
($x=0$): $x\to 0$, $\tau\to\tau$; the event rate is
$\gamma(\tau)$;
\\
{\sf (c)}\;the cleavage on the second ring of cleavage centers
($x=L$, $L$ is the distance between the rings of cleavage
centers): $x\to L$, $\tau\to\tau$; the event rate is
$\gamma(\tau-L)$.

In terms of these elementary events the possible scenarios of
transition $\tau\!\to\!\tau\!+\!1$ are

\noindent
{\sf (1)\;Elementary event (a)}. Its probability is
\[
P_1(x|\tau)=\left\{\begin{array}{cl}
v_x/(v_x+\gamma_\tau),& x\le L\,;\\[5pt]
v_x/(v_x+\gamma_\tau+\gamma_{\tau-L}),& x>L\,.
\end{array}\right.
\]
In this scenario, $x\to x+1$, and
\begin{equation}
w_1(x+1|\tau+1)=P_1(x|\tau)\,w(x|\tau)\,.
\label{eq_aux-1w}
\end{equation}
No peptides are generated;

\noindent
{\sf (2)\;Elementary event (b)}, which may not be followed by
anything but the strand shift by one AA (as there is nothing to be
cleaved). This scenario probability is
\[
P_2(x|\tau)=\left\{\begin{array}{cl}
\gamma_\tau/(v_x+\gamma_\tau),& x\le L\,;\\[5pt]
\gamma_\tau/(v_x+\gamma_\tau+\gamma_{\tau-L}),& x>L\,.
\end{array}\right.
\]
In this scenario, $x\to 1$, and
\begin{equation}
\textstyle
w_2(x|\tau+1)=\delta_{x,1}\sum_{x'=1}^\infty
P_2(x'|\tau)\,w(x'|\tau)\,.
\label{eq_aux-2w}
\end{equation}
The peptides cut out are
\begin{equation}
Q_2(\tau,\tau-x+1|\tau)=P_2(x|\tau)\,w(x|\tau)\,;
\label{eq_aux-2Q}
\end{equation}

\noindent
{\sf (3)\;Elementary event\,(c), which may be followed either by
strand shift\,(1) or by scenario\,(2)}. The probability of the
first stage (c) is
\[
P_c(x|\tau)=\left\{\begin{array}{cl}
0,& x\le L\,;\\[5pt]
\gamma_{\tau-L}/(v_x+\gamma_\tau+\gamma_{\tau-L}),& x>L\,.
\end{array}\right.
\]
After event (c), when $x\to L$, the number of the system states
generated is
\[
\textstyle
w_c(x|\tau)=\delta_{x,L}\sum_{x'=L+1}^\infty
P_c(x'|\tau)\,w(x'|\tau)\,,
\]
and the peptides cut out are
\[
Q_c(\tau-L,\tau-x+1|\tau)=P_c(x|\tau)\,w(x|\tau)\,.
\]
The subsequent events (1) or (2) should be considered as the
respective above mentioned scenarios starting with the
distribution $w_c(x|\tau)$, {\em i.e.},
\begin{eqnarray}
 &&\hspace{-17mm}
 w_{c1}(x|\tau+1)=P_1(L|\tau)\,w_c(x-1|\tau)
 =P_1(L|\tau)\;\delta_{x,L+1}\!\sum_{x'=L+1}^\infty\!P_c(x'|\tau)\,w(x'|\tau)\,,
\label{eq_aux-c1w}
\end{eqnarray}
\begin{eqnarray}
 &&\hspace{-17mm}
 Q_{c1}(\tau\!-\!L,\tau\!-\!x\!+\!1|\tau)
 =P_1(L|\tau)\,Q_c(\tau\!-\!L,\tau\!-\!x\!+\!1|\tau)
 =P_1(L|\tau)\,P_c(x|\tau)\,w(x|\tau)\,;
\label{eq_aux-c1Q}
\end{eqnarray}
\begin{eqnarray}
 &&\hspace{-17mm}
 w_{c2}(x|\tau+1)=\delta_{x,1}\sum_{x'=1}^\infty P_2(x'|\tau)\,w_c(x'|\tau)
 =\delta_{x,1}\;P_2(L|\tau)\!\sum_{x'=L+1}^\infty\!P_c(x'|\tau)\,w(x'|\tau)\,,
\label{eq_aux-c2w}
\end{eqnarray}
\begin{eqnarray}
 &&\hspace{-17mm}
 Q_{c2}(\tau\!-\!L,\tau\!-\!x\!+\!1|\tau)=P_2(L|\tau)\,Q_c(\tau\!-\!L,\tau\!-\!x\!+\!1|\tau)
 =P_2(L|\tau)\,P_c(x|\tau)\,w(x|\tau)\,,
\label{eq_aux-c2Q-1}
\end{eqnarray}
\begin{eqnarray}
 &&\hspace{-17mm}
 Q_{c2}(\tau,\tau-x+1|\tau)=P_2(x|\tau)\,w_c(x|\tau)
 =\delta_{x,L}\,P_2(L|\tau)\!\sum_{x'=L+1}^\infty\!P_c(x'|\tau)\,w(x'|\tau)\,.
\label{eq_aux-c2Q-2}
\end{eqnarray}

Collecting
equations~\eref{eq_aux-1w},\,\eref{eq_aux-2w},\,\eref{eq_aux-c1w},\,\eref{eq_aux-c2w},
one finds the {\em master equation}
\begin{eqnarray}
w(1|\tau+1)
 =\sum\limits_{x=1}^{L}\frac{\gamma_\tau\,w(x|\tau)}{v_x+\gamma_\tau}
 +\left(1+\frac{\gamma_{\tau-L}}{v_L+\gamma_\tau}\right)\hspace{-3pt}
 \sum\limits_{x=L+1}^{\infty}\frac{\gamma_\tau\,w(x|\tau)}{v_x+\gamma_\tau+\gamma_{\tau-L}};\label{eq01-1}\\[5pt]
w(L+1|\tau+1)
 =\frac{v_L}{v_L+\gamma_\tau}
 \biggl\lbrack w(L|\tau)
 +\sum\limits_{x=L+1}^{\infty}\frac{\gamma_{\tau-L}\,w(x|\tau)}{v_x+\gamma_\tau+\gamma_{\tau-L}}
 \biggr\rbrack;\label{eq01-2}\\[5pt]
w(x|\tau+1)
 =\frac{v_{x-1}\,w(x-1|\tau)}{v_{x-1}+\gamma_\tau+\Theta(x\!-\!L\!-\!1)\gamma_{\tau-L}}
 \quad\mbox{ for }\;x\ne 1,\;x\ne L+1.\label{eq01-3}
\end{eqnarray}
Here $x=1,2,3,...,M$ and $\tau=1,2,3,...,M-1$, where $M$ is the
length of the protein, and the Heaviside function
$\Theta(x\!<\!0)=0$, $\Theta(x\!\ge\!0)=1$.
Equations~\eref{eq01-1}--\eref{eq01-3} form a linear map
\begin{equation}
\textstyle
w(x|\tau+1)=\sum_{y=1}^\infty\mathcal{L}_{xy}(\tau)\,w(y|\tau).
\label{eq02}
\end{equation}

The whole contribution to the cleavage pattern
\begin{eqnarray}
 &&\hspace{-12mm}
 Q(\tau,\tau-x+1|\tau)=Q_2(\tau,\tau-x+1|\tau)
 +Q_{c2}(\tau,\tau-x+1|\tau)\nonumber\\[5pt]
 &&=\frac{\gamma_{\tau}\,w(x|\tau)}{v_x+\gamma_{\tau}+\Theta(x\!-\!L\!-\!1)\gamma_{\tau-L}}
 +\frac{\delta_{x,L}\,\gamma_{\tau}}{v_L+\gamma_{\tau}}
 \sum\limits_{x'=L+1}^M\frac{\gamma_{\tau-L}\,w(x'|\tau)}{v_{x'}+\gamma_{\tau}+\gamma_{\tau-L}}\,;
 \label{eq_aux-Q-1}\\[10pt]
 &&\hspace{-12mm}
 Q(\tau\!-\!L,\tau\!-\!L\!-\!x\!+\!1|\tau)=Q_{c1}(\tau\!-\!L,\tau\!-\!L\!-\!x\!+\!1|\tau)
 +Q_{c2}(\tau\!-\!L,\tau\!-\!L\!-\!x\!+\!1|\tau)\nonumber\\[5pt]
 &&\qquad=\frac{\gamma_{\tau-L}\,w(L+x|\tau)}{v_{L+x}+\gamma_{\tau}+\gamma_{\tau-L}}\,.
 \label{eq_aux-Q-2}
\end{eqnarray}
All the rest [not specified by
expressions~\eref{eq_aux-Q-1},\,\eref{eq_aux-Q-2}] elements
$Q(m,n|\tau)$ are zero. The expressions for digestion pattern
$Q(m,n)$ after the processing of a single protein molecule are
different for short polypeptides and long ones of a periodic AA
sequence.


\section{Short (25--50 AA) synthetic polypeptides}
\label{sec2}
First we consider degradation of {\em short} (25--50\,AA) {\em
synthetic polypeptide} (protein), the most common situation for
{\it in vitro} experiments. Here we start at $\tau=1$ with
$w(x|\tau=1)=\delta_{x,1}$ and iterate linear map~\eref{eq02} till
the last $\tau=M-1$. For a short polypeptide the releasing of the
last fragment from the chamber at the ``time moment'' $\tau=M$
should be additionally taken into account:
 $Q(M,M-x+1|M)\to Q(M,M-x+1|M)+w(x|M)$.
Hence, with $w(x|\tau)$ known for $\tau=1,2,...,M$, one may
evaluate digestion pattern $Q(m,n)$ from~\eref{eq_aux-Q-1}
and~\eref{eq_aux-Q-2},
\begin{eqnarray}
 &&\hspace{-10mm}
 Q(\tau_1,\tau_2)=Q(\tau_1,\tau_2|\tau_1)
 +\Theta(M\!-\!\tau_1\!-\!L)\,Q(\tau_1,\tau_2|\tau_1+L)
 \nonumber\\[5pt]
 &&
 =\delta_{\tau_1,M}w(\tau_1+L-\tau_2+1|M)
 +\frac{\gamma_{\tau_1}\,w(\tau_1-\tau_2+1|\tau_1)}
 {v_{\tau_1-\tau_2+1}+\gamma_{\tau_1}+\Theta(\tau_1\!-\!\tau_2\!-\!L)\gamma_{\tau_1-L}}\nonumber\\
 &&\qquad
 +\frac{\delta_{\tau_1-\tau_2+1,L}\,\gamma_{\tau_1}}{v_L+\gamma_{\tau_1}}
 \!\sum_{x=L+1}^M\frac{\gamma_{\tau_1-L}\,w(x|\tau_1)}{v_x+\gamma_{\tau_1}+\gamma_{\tau_1-L}}
 \nonumber\\
 &&\qquad\qquad
 +\Theta(M\!-\!\tau_1\!-\!L)
 \frac{\gamma_{\tau_1}\,w(\tau_1+L-\tau_2+1|\tau_1+L)}
 {v_{\tau_1+L-\tau_2+1}+\gamma_{\tau_1+L}+\gamma_{\tau_1}},
 \label{eq03}
\end{eqnarray}
here $1\le\tau_2\le\tau_1\le M$. Since the protein may be cleaved
starting both from the C- and from the N-terminal, the final
digestion pattern is given by
\begin{equation}
 Q_\mathrm{fin}(\tau_1,\tau_2)=P_\Nt\,Q_\Nt(\tau_1,\tau_2)
 +P_\Ct\,Q_\Ct(M-\tau_2+1,M-\tau_1+1)\,.\label{eq04}
\end{equation}
The subscripts indicate which terminal goes first, $P_\Nt$ and
$P_\Ct=1-P_\Nt$ are the probabilities of the degradation starting
from the corresponding end. Generally, $v_\Nt(x)$ and $v_\Ct(x)$
may be slightly different, but here we neglect this difference.
Note that a fragment length distribution $S(x)$ (often used in the
literature~\cite{1998_Kisselev_jbc}) is then the convolution
\begin{equation}
\textstyle
S(x)=\sum_{\tau=x}^{M}Q(\tau,\tau-x+1)\,.
\label{eq_S}
\end{equation}

\begin{figure}[!t]
\center{
  \includegraphics[width=0.96\textwidth]%
 {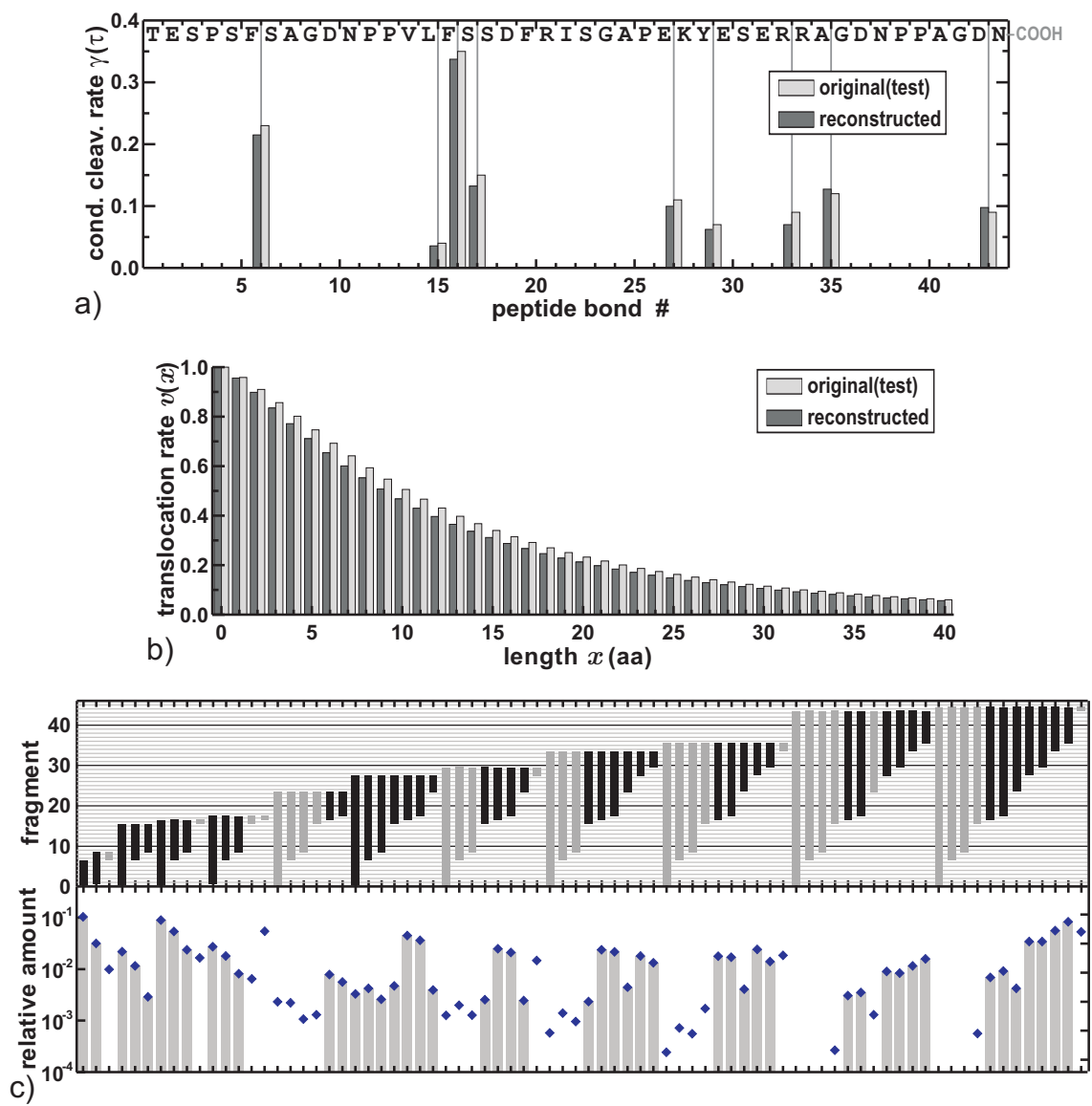}
}
  \caption{Test --- Reconstruction of translocation rate function
$v(x)$ and conditional cleavage rates $\gamma(\tau)$ for the 44mer
peptide Kloe~316~\cite{2006_Mishto,2008_Mishto_Zaikin_proteamalg}
[but with roughly estimated authentic (original) values of
$\gamma(\tau)$], which is the subsequence 543--586\,AA of human
myelin associated glycoprotein. a)~the conditional cleavage rates
and the AA sequence; b)~the translocation rate function; c)~the
upper plot presents the set of digestion fragments (black bars:
fragments utilized for the reconstruction, grey bars: not
utilized), and the lower plot presents the amount of the
corresponding fragment (diamonds: the reconstructed values
$Q_\mathrm{fin}$, grey bars: the values of $\widetilde{Q}$
utilized for the reconstruction).
}
  \label{fig2}
\end{figure}

Digestion pattern $Q_\mathrm{fin}(\tau_1,\tau_2)$ is a functional
of TRF $v(x)$ and CCR $\gamma(\tau)$. Utilizing MS data on the
digestion pattern, one can determine nonzero values of
$\gamma(\tau)$ ({\it i.e.}\ positions of possible cleavage) and
minimize the mismatch between $Q_\mathrm{fin}(\tau_1,\tau_2)$ and
MS data $\widetilde{Q}(\tau_1,\tau_2)$ over $v(x)$, the nonzero
values of $\gamma(\tau)$, and $P_\Nt$ in order to {\it
reconstruct} them. Expecting the function $v(x)$ to be smooth, we
parameterize appropriate approximate functions as
\begin{equation}
v_\mathrm{app}(x)=v_0\e^{-\frac{A_2^2}{\sqrt{A_1^2+x}}+\frac{A_2^2}{|A_1|}-A_3^2(\sqrt{A_1^2+x}-|A_1|)}.
\label{eq05}
\end{equation}
Note, $v(x)$ and $\gamma(\tau)$ are defined up to a constant
multiplier, which should be determined from the degradation rate
in real time, but not from the digestion pattern.


In order to verify the robustness of the reconstruction procedure,
numerous tests have been performed. A typical test presented in
\fref{fig2} has been performed in 4 steps:

\noindent
{\sf (1)}\;For given $v(x)$ [not generic for $v_\mathrm{app}$,
{\it i.e.}, the used function $v(x)$ cannot be perfectly fitted
with expression~\eref{eq05}] and $\gamma(\tau)$ digestion pattern
$Q(\tau_1,\tau_2)$ has been evaluated.

\noindent
{\sf (2)}\;The result has been perturbed by the noise,
 $\widetilde{Q}_{\tau_1\tau_2}=Q_{\tau_1\tau_2}
 +10^{-4}R_{\tau_1,\tau_2}\sqrt{Q_{\tau_1,\tau_2}}$,
where $R_{\tau_1,\tau_2}$ are independent random numbers uniformly
distributed in $[-1,1]$.

\noindent
{\sf (3)}\;We have omitted the information about fragments, which
relative amount is less than $5\cdot10^{-3}$, and 1mer and 2mer
fragments as being hardly detectable in experiments (one cannot
distinguish identical AAs cut out from different parts of the
polypeptide~\cite{2001_Kohler}).

\noindent
{\sf (4)}\;Resulting $\widetilde{Q}_{\tau_1\tau_2}$ has been used
for the reconstruction of $v(x)$ and $\gamma(\tau)$.

\noindent
The original and reconstructed data for $\gamma(\tau)$
(\fref{fig2}a) and $v(x)$ (\fref{fig2}b) are in a very good
agreement. The reconstructed $P_\Nt=0.52$ against original
$P_\Nt=0.50$\,.

Unfortunately, the data available in the literature are mainly too
much incomplete (a lot of fragments are not accounted) and not
enough precise for a truly reliable
reconstruction~\cite{2008_Goldobin_etal} (the initial solutions
used for experiments quite frequently contain not only the
polypeptide to be digested but also a certain amount of its
fragments, the first measurement of the proportions of the
solution is performed to late, when considerably more than 5\% of
the initial substrate has been degraded and one may not neglect
reentries of the digestion fragments into the proteasome, {\it
etc.}).

Thus, we should note the limitations of the suggested
reconstruction method:

\noindent
$\bullet$\;The reconstruction procedure for short polypeptides is
very sensitive to  measurement inaccuracy.

\noindent
$\bullet$\;For some polypeptides the procedure fails. This may
happen due to a specific arrangement of cleavage positions, when
different TRFs $v(x)$ provides almost identical digestion
patterns.

\noindent
$\bullet$\;Though the whole information on $Q(\tau_1,\tau_2)$ is
not needed, the number of nonzero values of $Q(\tau_1,\tau_2)$
required for a reliable (tolerant to noise) reconstruction is at
least the twice number of reconstructed parameters, {\it i.e.}\
 $2\times([
\footnotesize\mbox{\sf number of positions of potential cleavage}]
 +[\footnotesize\mbox{\sf number of parameters of }v_\mathrm{app}]
 +1)$.
For instance, for Kloe~258 in~\cite{2008_Goldobin_etal} the number
of trustworthy and utilized values of
$\widetilde{Q}(\tau_1,\tau_2)$ is $19$ instead of the required
$2\times(10+3+1)=28$, it is a bit greater than the number of the
unknown parameters, {\it i.e.}, 14. Hence, more accurate and
comprehensive MS data on the digestion pattern are required.

\noindent
$\bullet$\;For short polypeptides the finishing stage of the
degradation is relatively important, while in this stage the
translocation rate is affected by the edge effects (the backward
end of the polypeptide gets inside the proteasome chamber) and is
not the same as for the remainder of the polypeptide.


\section{Long synthetic polypeptides of a periodic amino acid sequence}
\label{sec3}
While a more comprehensive acquisition of data on digestion
fragments and enhancement of experimental techniques for short
polypeptides are up to experimentalists  we propose experimental
set-up which allows overcoming all the limitations mentioned above
and is expected to be realizable. For this a {\em long synthetic
polypeptide with a $T$-periodic AA sequence}:
$\gamma(\tau)=\gamma(\tau+T)$ should be digested.  Here ``long''
means one may neglect the peculiarities of the starting and
finishing stages of the degradation, and $M\gg T$.

For the given direction of the degradation, {\it e.g.}, starting
with the N-terminal, we are looking for the establishing
$T$-periodic in $\tau$ solution
$w_{\Nt,T}(x|\tau)=w_{\Nt,T}(x|\tau-T)$ to equation~\eref{eq02}.
The fragment $(n,m)$ is identical to the one
$(n\!+\!kT,m\!+\!kT)$, where $k$ is integer; therefore
$Q_\Nt(m,n)$ may be chosen to make contribution to
$Q_\Nt(m-n+{(n\;\mathrm{mod}\;T)},\,{n\;\mathrm{mod}\;T})$. The
amount of fragments grows almost linearly with ``time'' $\tau$ as
the polypeptide being processed. Hence, for the digestion pattern
one finds
\begin{eqnarray}
 \hspace{-10mm}Q_{\Nt,T}(\tau_1,\tau_2)
 \equiv\lim\limits_{\tau\to\infty}\frac{1}{\tau}\sum_{\tau'=1}^{\tau}Q_\Nt(\tau_1,\tau_2|\tau')
 =\frac{1}{T}\sum_{\tau'=1}^TQ_{\Nt,T}(\tau_1,\tau_2|\tau')
\nonumber\\
 =\frac{1}{T}\Bigg[\frac{\gamma_{\tau_1}\,w_{\Nt,T}(\tau_1-\tau_2+1|\tau_1)}
 {v_{\tau_1-\tau_2+1}+\gamma_{\tau_1}+\Theta(\tau_1\!-\!\tau_2\!-\!L)\gamma_{\tau_1-L}}\nonumber\\
 \hspace{-10mm}
 {}+\frac{\delta_{\tau_1-\tau_2+1,L}\,\gamma_{\tau_1}}{v_L+\gamma_{\tau_1}}
 \!\sum_{x=L+1}^\infty\frac{\gamma_{\tau_1-L}\,w_{\Nt,T}(x|\tau_1)}{v_x+\gamma_{\tau_1}+\gamma_{\tau_1-L}}
 +\frac{\gamma_{\tau_1}\,w_{\Nt,T}(\tau_1+L-\tau_2+1|\tau_1+L)}
 {v_{\tau_1+L-\tau_2+1}+\gamma_{\tau_1+L}+\gamma_{\tau_1}}\Bigg]
\label{eq06}
\end{eqnarray}
(here $1\le\tau_2\le T$ and $\tau_1\ge\tau_2$).

To treat the degradation process starting with the C-terminal, one
has (i)~to perform the transformation
$\gamma(\tau)\to\gamma(T-\tau)$, (ii)~iterate linear
map~\eref{eq02} with the new $\gamma(\tau)$ like for the N-case,
but assuming $Q_\Ct(m,n|\tau)$ to make contribution to
$Q_\Ct({m\;\mathrm{mod}\;T},\,n-m+{(m\;\mathrm{mod}\;T)})$.
Unlike~\eref{eq04}, the final result is
\[
 Q_\mathrm{fin}(\tau_1,\tau_2)
 =P_\Nt Q_{\Nt,T}(\tau_1,\tau_2)
 +P_\Ct Q_{\Ct,T}(T\!-\!\tau_2,T\!-\!\tau_1).
\]

\begin{figure}[!t]
\hspace{10mm}
\center{
  \includegraphics[width=0.96\textwidth]%
 {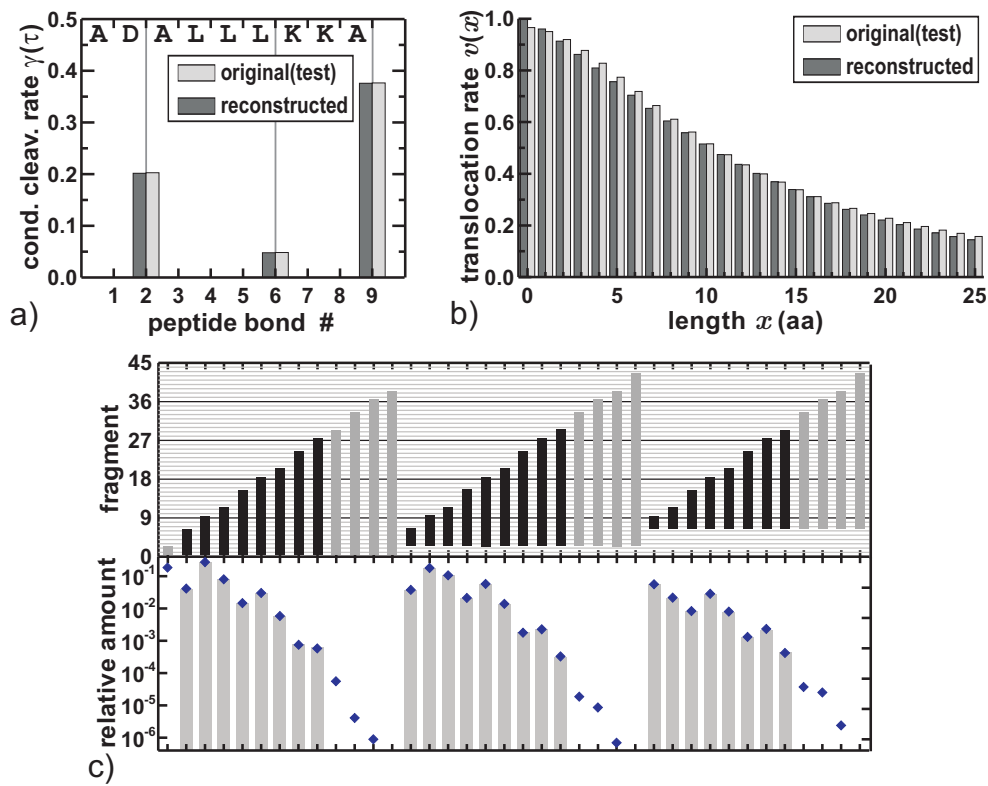}
}
  \caption{
Test --- Reconstruction of translocation rate function $v(x)$ and
conditional cleavage rates $\gamma(\tau)$ for a 9-periodic
polypeptide with the cleavage positions 2, 6, 9. (For description
see caption to \fref{fig2}.)
}
  \label{fig3}
\end{figure}

Matching $Q_\mathrm{fin}(m,n)$ to the MS data one can reconstruct
$v(\tau)$, $\gamma(\tau)$, and $P_\Nt$. For a test we have made
use of the cleavage map of the digestion of yeast enolise-1 by
human erythrocyte proteasome~\cite{2001_Nussbaum}. Looking at its
subsequence 331--348\,AA
\[
 ...|
\mbox{\tt ATAIE}\underline{\mbox{\tt KKA}|\mbox{\tt AD}|\mbox{\tt ALLL}}|\mbox{\tt KV}|\mbox{\tt NQ}|...
\mbox{\footnotesize --COOH}
\]
(vertical stripes mark the positions of experimentally observed
cleavages), one may expect the case, where the underlined
subsequence is followed not by $\mbox{\tt KV}$, but by $\mbox{\tt
KKA}...$, and the periodic sequence is
\[
\mbox{\tt AD}|\mbox{\tt ALLL}|\mbox{\tt KKA}|\dots|
\mbox{\tt AD}|\mbox{\tt ALLL}|\mbox{\tt KKA}|\dots|
\mbox{\tt AD}|\mbox{\tt ALLL}|\mbox{\tt KKA}\mbox{\footnotesize --COOH}\,,
\]
to be realizable. For such a sequence a test like the one in
\fref{fig2} (but with much stronger dithering:
$\widetilde{Q}_{\tau_1\tau_2}=Q_{\tau_1\tau_2}
 +2\cdot 10^{-3}R_{\tau_1,\tau_2}\sqrt{Q_{\tau_1,\tau_2}}$)
is presented in \fref{fig3}. Due to the small number of unknown
parameters the reconstruction procedure is rather tolerant to
measurement inaccuracy and does not require information on a large
number of digestion fragments (the most easily detectable
fragments are enough).


\section{Long natural proteins}
\label{sec4}
The case of a most immediate interest is the digestion of long
natural proteins (over about 300\,AA) because it concerns the {\it
in vivo} proteasomal activity. A direct implementation of the
procedure developed for short polypeptides is hardly possible
here, as in the course of matching $Q(\tau_1,\tau_2)$ to the MS
data, one has to perform a minimization over an enormous number of
parameters. However, for long non-periodic proteins, one may
assume $\gamma(\tau)$ to be a random process in order to evaluate
some observable statistical properties like the fragment length
distribution (FLD) of the digestion products, {\it i.e.} $S(x)$
[see equation~\eref{eq_S}].

For this random process we adopt the following:
\\
$\bullet$\;the neighbor values $\gamma(\tau)$ and
$\gamma(\tau\!+\!1)$ are mutually independent (what does not
necessarily mean that CCR $\gamma(\tau)$ is independent of
neighbor AAs);
\\
$\bullet$\;$\gamma(\tau)$ is zero with a certain probability $q$,
and has a finite probability density $g(\gamma)$ otherwise.

The normalized mean FLD
 $\mathcal{S}(x)\equiv\la S(x)\ra/\sum_{x'=1}^\infty\la S(x')\ra$
may be evaluated either via the plain iterating
of~\eref{eq02}--\eref{eq04} with noise $\gamma(\tau)$ over a large
interval of $\tau$ or via the direct simulation of the system with
a Gillespie algorithm ({\it e.g.}, see~\cite{2006_Zaikin_Goldobin}).
However, the calculation procedure may be considerably
facilitated. For this purpose, let us average~\eref{eq02} over
realizations of $\gamma(\tau)$,
\begin{equation}
\textstyle
\la w(x|\tau+1)\ra_\gamma
 =\sum_{y=1}^{\infty}\la\mathcal{L}_{xy}(\tau)\,w(y|\tau)\ra_\gamma\,.
\label{eq-noise-01}
\end{equation}
Noteworthy, $w(x|\tau)$ depends on $\gamma(\tau-1)$ and the
preceding values of $\gamma$ but is independent of $\gamma(\tau)$.
Moreover, the impact of preceding values of $\gamma$ decays in the
course of the processing of the protein, and one may neglect the
correlation between $w(x|\tau)$ and $\gamma(\tau-L)$ which are
mutually distant in $\tau$. Thus, $w(x|\tau)$ is independent of
$\gamma(\tau)$ and $\gamma(\tau-L)$, which are involved in
$\mathcal{L}_{xy}(\tau)$, and~\eref{eq-noise-01} yields
\begin{equation}
\textstyle
\la w(x|\tau+1)\ra_\gamma
 \approx\sum_{y=1}^{\infty}\la\mathcal{L}_{xy}(\tau)\ra_{\gamma_\tau\gamma_{\tau-L}}\la w(y|\tau)\ra_\gamma\,;
\label{eq-noise-02}
\end{equation}
from~\eref{eq_aux-Q-1},\,\eref{eq_aux-Q-2},\,\eref{eq_S},
\begin{eqnarray}
\hspace{-10mm}
\la S(x|\tau+1)\ra_\gamma=\la S(x|\tau)\ra_\gamma
 +\lla\displaystyle\frac{\gamma_\tau}
 {v_x+\gamma_\tau+\Theta(x\!-\!L\!-\!1)\gamma_{\tau-L}}\rra_{\gamma_\tau\gamma_{\tau-L}}\hspace{-5pt}
 \lla w(x|\tau)\rra_\gamma
 \nonumber\\[5pt]
 {}+\lla\displaystyle\frac{\gamma_{\tau-L}}{v_{L+x}+\gamma_\tau+\gamma_{\tau-L}}\rra_{\gamma_\tau\gamma_{\tau-L}}\hspace{-5pt}
 \lla w(L+x|\tau)\rra_\gamma
 \nonumber\\[3pt]
 \qquad
 {}+\delta_{x,L}\!\sum\limits_{x'=L+1}^\infty\!\lla\frac{\gamma_{\tau}}{v_L+\gamma_{\tau}}\cdot
 \frac{\gamma_{\tau-L}}{v_{x'}+\gamma_{\tau}+\gamma_{\tau-L}}\rra_{\gamma_\tau\gamma_{\tau-L}}\hspace{-5pt}
 \lla w(x'|\tau)\rra_\gamma,
\label{eq-noise-03}
\end{eqnarray}
where
\[
\begin{array}{r}
 \hspace{-15mm}
 \la f(\gamma_1,\gamma_2)\ra_{\gamma_1\gamma_2}
 \equiv q^2f(0,0)+q(1-q)\int_0^\infty
 g(\gamma)[f(0,\gamma)+f(\gamma,0)]\rmd\gamma\qquad\qquad\\[5pt]
 {}+(1-q)^2\int_0^\infty\rmd\gamma_1\int_0^\infty\rmd\gamma_2\,g(\gamma_1)\,g(\gamma_2)\,f(\gamma_1,\gamma_2)\,.
\end{array}
\]
The FLD observed in experiments is $\mathcal{S}(x)$ corresponding
to the establishing steady solution
 $\la w(x|\infty)\ra$
to linear map~\eref{eq-noise-02}.

Noteworthy, with the additional approximation
\[
 \la\mathcal{L}_{xy}(\gamma_\tau,\gamma_{\tau-L})\ra_{\gamma_\tau\gamma_{\tau-L}}
 \approx\mathcal{L}_{xy}(\la\gamma\ra,\la\gamma\ra)\,,
\]
one may obtain an implicit recursive formula for establishing
 $\la w(x|\tau)\ra$
from~\eref{eq-noise-02},
\begin{equation}
\la w(x+1|\infty)\ra
 =\frac{(1+\delta_{x,L})\,v_x}{v_x+(1+\Theta(x-L))\la\gamma\ra}\la w(x|\infty)\ra\,,
\label{eq-noise-04}
\end{equation}
and find FLD $\mathcal{S}(x)$ from~\eref{eq-noise-03},
\begin{equation}
\mathcal{S}(x)=
 \frac{\displaystyle
 \frac{(1+\delta_{x,L})\la\gamma\ra\;\la w(x|\infty)\ra}{v_x+(1+\Theta(x-L))\la\gamma\ra}
 +\frac{\la\gamma\ra\;\la w(L+x|\infty)\ra}{v_{L+x}+2\la\gamma\ra}}
 {\displaystyle
 \la w(1|\infty)\ra+\frac{\la w(L+1|\infty)\ra}{2}}\,.
\label{eq-noise-05}
\end{equation}
Remarkably, in the quasi-continuous limit (which is valid when
$v(x)$ is a ``slow'' function of $x$), the last expressions
provide (cf.\,\cite{2006_Zaikin_Goldobin})
\[
\la w(x|\infty)\ra
 =(1+\Theta(x-L))\la w(0|\infty)\ra\,\e^{-\int\limits_0^x\frac{(1+\Theta(x-L))\,\la\gamma\ra}{v(x')}\rmd x'},
\]
\[
\mathcal{S}(x)=
 \la\gamma\ra\;\frac{\displaystyle
 \frac{\e^{-\int\limits_0^x\frac{(1+\Theta(x-L))\,\la\gamma\ra}{v(x')}\rmd x'}}{v(x)}
 +\frac{\e^{-\int\limits_0^{L+x}\frac{(1+\Theta(x-L))\,\la\gamma\ra}{v(x')}\rmd x'}}{v(L+x)}}
 {\displaystyle
 1+\e^{-\int\limits_0^L\frac{\la\gamma\ra}{v(x')}\rmd x'}}\,.
\]

\begin{figure}[!t]
\hspace{10mm}
\center{
  \includegraphics[width=0.98\textwidth]%
 {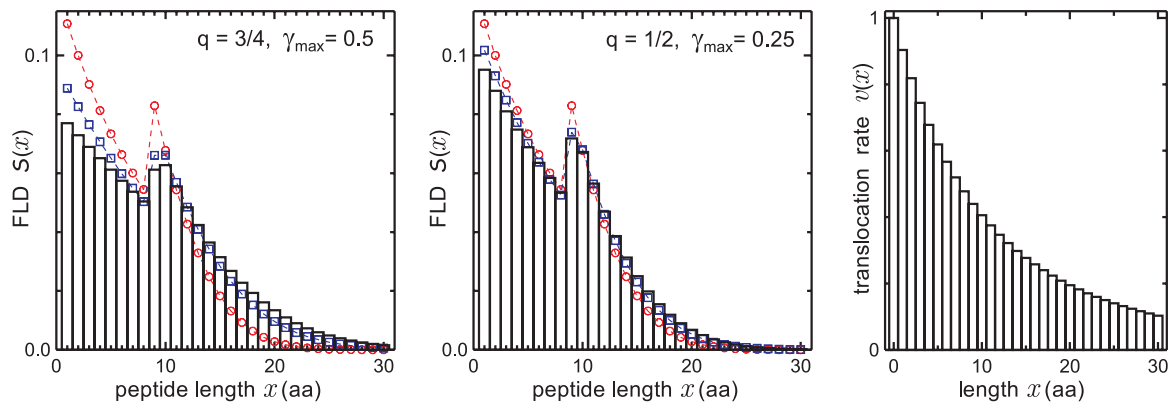}
}
  \caption{
Samples of fragment length distribution $\mathcal{S}(x)$ (FLD) for
the degradation of a long natural protein under the assumption,
that conditional cleavage rate $\gamma(\tau)$ may be treated as a
random process. The fraction $q$ of nonscissile peptide bonds is
indicated in the plots, nonzero values of $\gamma(\tau)$ are
uniformly distributed in $[0,\gamma_\mathrm{max}]$, $L=9$, the
adopted translocation rate function $v(x)$ is plotted in the right
plot. In two left plots, bars: results of the direct simulation
with a Gillespie algorithm, squares: the
approximation~\eref{eq-noise-02},\,\eref{eq-noise-03}, circles:
the approximation~\eref{eq-noise-04},\,\eref{eq-noise-05} with
$\la\gamma\ra=(1-q)\gamma_\mathrm{max}/2$.
}
  \label{fig4}
\end{figure}

In \fref{fig4}, one may see, that the both above mentioned
approximations become more accurate as $q$ decreases. However, for
realistic value $q\approx3/4$ which is suggested by experimental
cleavage maps (see \fref{fig2}, where the sites of a potential
cleavage are taken from experimental data), the
approximation~\eref{eq-noise-02},\,\eref{eq-noise-03} works
considerably better than the
one~\eref{eq-noise-04},\,\eref{eq-noise-05}. Remarkably, as $q$
increases with $\la\gamma\ra$ kept fixed, the local maximum near
$x=L$ shifts from $x=L$ to higher values of fragment length $x$
and the cutting-out of longer peptides becomes more probable. The
existence of this maximum at $L\approx8-10$\,AA deserves especial
attention because the epitopes,  involved in the functioning of
the immune system and bound to MHC~I molecules, have exactly such
length ~\cite{1991_Falk-1991_Madden}.

The important limitation of this method is related to the
reconstruction of $v(x)$ for 1mer and 2mer peptides. These
peptides are hardly detectable in experiments and, therefore,
experimental $\mathcal{S}(x)$ is not determined for $x=1,\,2$, and
one cannot reconstruct the respective values of $v(x)$. Note, for
methods suggested in sections~\ref{sec2} and~\ref{sec3} this
limitation does not occur because, {\it e.g.}, for the subsequence
{\tt |F|S|SDFRISGAPE|} in
\fref{fig2}, the information on $v(1)$ is reflected in the
difference between the readily measurable amounts of generated
peptides {\tt |S|SDF...|} and {\tt |SDF...|}, while for long
natural proteins we lose the individual information on each
specific peptide cut out.

\section{Conclusion}
\label{concl}
In this paper we have discussed a model of the degradation of
proteins by the proteasome which allows one to {\em reconstruct}
the proteasomal translocation function and the cleavage
specificity inherent to the amino acid sequence and not affected
by proteasomal transport properties. With these properties
determined, one can comprehensively predict digestion patterns of
new proteins. The model is relevant for a broad variety of
hypothetically possible translocation
mechanisms~\cite{2005_Zaikin_proteasome,2001_Reimann}. We have
mathematically elaborated this model for the cases of
(i)~relatively short (25--50mers) synthetic polypeptides as the
most common case for {\it in vitro} experiments, (ii)~long
periodic polypeptides (proposed experiments with such polypeptides
are very promising for reverse engineering), and (iii)~long
natural proteins.

In~\cite{2006_Zaikin_Goldobin}, we have already discussed how
peculiarities of the translocation function may lead to the
multimodality of the fragment length distribution even for
$\gamma(\tau)=const$. Here we have shown that the amount of each
digestion fragment is not only determined by the cleavage map
[specifically, conditional cleavage rate $\gamma(\tau)$] of the
substrate but is also crucially affected by nonuniformity of the
translocation rate. The results of implementation of the developed
theory for processing experimental data on digestion patterns for
different proteasome species under different conditions can give
insight into the nature of the protein translocation mechanism
inside the proteasome. They can as well elucidate the unanswered
question whether there is some preference for starting the
degradation with the N- or C-terminal of the protein, and how this
preference is affected by regulatory complexes. Hopefully,
theoretical results will stimulate new experiments as suggested in
this paper for the case of a periodic polypeptide.

\ack{
We are thankful to Susanne Witt and Michele Mishto for useful
discussions. The work has been supported by grants of the
VW-Stiftung, the BRHE program, and the Foundation ``Perm
Hydrodynamics''.}

\end{document}